\documentclass[notoc,12pt]{JHEP3}
\pdfoutput=1
\usepackage{amsfonts}
\usepackage{amscd}
\usepackage{amssymb}
\usepackage{amsmath}
\usepackage{epsfig}
\usepackage{latexsym}
\usepackage{mathtools}

\def \be  {\begin{equation}}
\def \ee  {\end{equation}}
\def \ba  {\begin{eqnarray}}
\def \ea  {\end{eqnarray}}
\def \bb  {\begin{thebibliography}}
\def \eb  {\end{thebibliography}}
\def \lab #1 {\label{#1}}
\newcommand\PT{\mathbb{PT}}
\newcommand\cA{\mathcal{A}}

\newcommand\cG{\mathcal{G}}

\newcommand\cK{\mathcal{K}}

\newcommand\cO{\mathcal{O}}

\newcommand\cW{\mathcal{W}}

\newcommand\cN{\mathcal{N}}

\newcommand\M {\mathbb{M }}

\newcommand\CP {\mathbb{CP}}

\newcommand\rd{\mathrm{d}}

\newcommand\im{\mathrm{i}}
\newcommand\la{\langle}
\newcommand\ra{\rangle}
\newcommand\del{\partial}
\newcommand\delbar{\bar{\partial}}

\title{A Proof of the Correlation Function /  Supersymmetric Wilson Loop Correspondence}

\author{{Tim Adamo$^*$, Mathew Bullimore$^\dagger$, Lionel Mason$^*$ and David Skinner$^\ddagger$}\\
	{\it $^*$The Mathematical Institute,\\
	24-29 St.~Giles', Oxford OX1 3LB, United Kingdom\\
	
	$^\dagger$Rudolf Peierls Centre for Theoretical Physics,\\
	1 Keble Road, Oxford OX1 3NP, United Kingdom\\
	
	$^\ddagger$Perimeter Institute for Theoretical Physics,\\ 
	31 Caroline St., Waterloo, ON, N2L 2Y5, Canada}}

\abstract{We prove that in the limit when its insertion points become pairwise null-separated, the ratio of certain $n$-point 
correlation functions in $\cN=4$ SYM is equal to a supersymmetric Wilson loop on twistor space, acting in the adjoint representation. In the planar limit, each of these objects reduces to the square of the complete $n$-particle planar superamplitude. Our proof is at the level of the integrand.}

\begin{document}
\maketitle


\section{Introduction}
\label{sec:intro}

Over the past few years, there has been much interest~\cite{Alday:2007hr,Drummond:2007aua,Brandhuber:2007yx,Drummond:2008aq,Anastasiou:2009kna,DelDuca:2010zg,Goncharov:2010jf,Gorsky:2009dr,Alday:2010vh,Gaiotto:2011dt} in a correspondence 
between planar scattering amplitudes in $\cN=4$ Super Yang-Mills (SYM) and the planar limit of the correlation function of a Wilson loop 
in the fundamental representation, stretched around a piecewise null polygonal contour in Minkowksi space. 

In~\cite{Mason:2010yk,Bullimore:2011ni}, we showed that the planar integrand of a supersymmetric Wilson loop in twistor space, taken 
in the fundamental representation, agrees with the integrand of the planar scattering amplitude -- including all N$^k$MHV partial 
amplitudes to all orders in the 't Hooft coupling, divided by an overall factor of the MHV tree amplitude. In~\cite{Mason:2010yk}, this 
correspondence was  conjectured, with the supporting evidence that the Feynman diagrams of the Wilson loop correlator (computed 
using the twistor action~\cite{Boels:2006ir} for $\cN=4$ SYM in axial gauge) explicitly reproduce the standard MHV diagram expansion 
for scattering amplitudes~\cite{Cachazo:2004kj,Elvang:2008vz,Bullimore:2010pj,Bullimore:2010dz}. In~\cite{Bullimore:2011ni} the 
conjecture was proved by using a twistor version of the Migdal-Makeenko equations to show that the Wilson loop planar integrand obeys 
the all-loop extension of the BCFW recursion relation for the integrand of the planar scattering amplitude~\cite{Arkani-Hamed:2010kv}. 
Closely related work was performed by Caron-Huot~\cite{Caron-Huot:2010ek} from the space-time perspective.

\medskip

In a very interesting series of papers~\cite{Alday:2010zy,Eden:2010zz,Eden:2010ce}  Alday, Eden, Korchemsky, Maldacena and 
Sokatchev have suggested a remarkable extension of the correspondence between Wilson loops and scattering amplitudes to include 
correlation functions. More specifically, these authors conjecture that
\be
	\lim_{x_{i,i+1}^2 \to 0}\, \frac{G(x_1,\cdots,x_n)}{G^{(0)}(x_1,\cdots,x_n)}
	= \la W_{\rm adj}[\gamma]\ra
\label{conj}
\ee
where $W_{\rm adj}[\gamma]$ is a (non-supersymmetric) Wilson loop in the adjoint representation (normalised so that its leading term is 1), taken along the null polygonal contour
\be
	\gamma = (x_1x_2)\cup(x_2x_3)\cup\cdots\cup(x_n x_1)
\label{polygon}
\ee
that appears in the limit when adjacently labelled insertion points in the correlator 
\be
	G(x_1,\ldots,x_n) \equiv \big\la\prod_{i=1}^n \cO(x_i)\big\ra
\label{Gdef}
\ee
become null separated. $\eqref{Gdef}$ is the $n$-point correlator of a local, gauge invariant operator such as the 1/2 BPS operator
\be
	\cO_{abcd}={\rm Tr}( \Phi_{ab}\Phi_{cd})-\frac{1}{12}\epsilon_{abcd}{\rm Tr}(\overline{\Phi}^{ef}\Phi_{ef})
\label{BPS}
\ee
and in~\eqref{conj},  $G^{(0)}(x_1,\ldots,x_n)$ is the same correlator at tree level only. The ratio $G/G^{(0)}$ and the Wilson loop 
expectation value are each divergent in the limit $x_{i,i+1}^2\to0$, so to compare both sides one must either work with a regularised 
scheme such as dimensional regularisation, or else carefully consider the limiting behaviour of the correlator and limiting behaviour of a 
Wilson loop on a nearby, non-null curve. These approaches were each considered in~\cite{Alday:2010zy}.

Importantly however,~\cite{Eden:2010ce} further conjectured that, as for the scattering amplitude/Wilson loop correspondence, this new 
correspondence should hold already at the level of the {\it integrand}\footnote{The notion of the {\it integrand} was introduced 
in~\cite{Arkani-Hamed:2010kv} in the context of planar scattering amplitudes. It does not make sense for arbitrary $n$-point correlation 
functions, as there is no way to compare different contractions between gauge invariant operators. However, it is meaningful 
in~\eqref{conj} precisely because only one way of contracting the fields survives in the limit of null separation.}. The integrands may be 
viewed as the sum of all Feynman diagrams that contribute to the correlator {\it before} the locations of any interaction vertices -- {\it i.e.}, 
the loop integrals -- are performed. The integrand is thus a rational function of the locations of both the operator insertions and any 
interaction vertices.

\medskip

The contour $\gamma$ that appears in the limit $x_{i,i+1}^2\to0$ is the same piecewise null curve that plays an important role in 
scattering amplitudes, so this story should be naturally related to the scattering amplitude / fundamental Wilson loop correspondence. 
In the planar limit, the correlation function of an adjoint Wilson loop reduces to the product of correlators of a fundamental and anti-
fundamental Wilson loop. In a CPT invariant theory these two factors are equal, so if the above conjectured relation between correlators 
and adjoint Wilson loops could be proved, it would follow immediately from the proof of~\cite{Bullimore:2011ni} that -- in the planar limit -- 
the ratio of correlation functions in~\eqref{conj} is also given by the square $(A_{\rm MHV}/A_{\rm MHV}^{(0)})^2$ of the ratio of the 
all-loop {\it MHV} amplitude to the MHV tree.

\medskip

The aim of this paper is to prove the above conjectures to all orders in g$^2$ {\it at the level of the integrand}, using twistor theory. As for 
the amplitude / Wilson loop correspondence, one of the main advantages of performing the calculation in twistor space is that it the 
supersymmetric extension may be handled straightforwardly. In twistor space the fundamental $\cN=4$ SYM multiplet may be described 
{\it off-shell} by the superfield\footnote{In this paper we use $(Z^\alpha,\chi^a)$ to denote homogeneous coordinates on $\CP^{3|4}$. We 
shall sometimes write $Z^\alpha= (\lambda_A,\mu^{A'})$ in terms of two 2-component spinors $\mu$ and $\lambda$.  In Penrose 
conventions, this would be a dual twistor.}
\be
	\cA(Z,\bar{Z},\chi) =
	a(Z,\bar Z) + \chi^a\gamma_a(Z,\bar Z) + \,\cdots\, + \chi^1\chi^2\chi^3\chi^4 g(Z,\bar Z)
\label{supermultiplet}
\ee
introduced by Ferber in~\cite{Ferber:1977qx} and exploited to great effect by Witten in~\cite{Witten:2003nn}. The coefficient of $(\chi)^m$ 
in~\eqref{supermultiplet} is a smooth (0,1)-form on $\CP^{3|4}$, homogeneous of degree $-m$. The linearised field equations in twistor 
space state that $\cA$ is holomorphic in $Z$, whereupon the component fields correspond to the linearised on-shell $\cN=4$ 
supermultiplet in space-time. However,  we repeat that $\cA(Z,\bar Z,\chi)$ is first and foremost an off-shell field in twistor space. We shall 
see that writing twistor expressions in terms of this superfield provides a natural supersymmetrization of both sides of the 
correspondence~\eqref{conj}, such that the ratio of $n$-point correlators of certain supermultiplets corresponds to a certain 
supersymmetric Wilson loop. This supersymmetric twistor Wilson loop is exactly the same object that appeared 
in~\cite{Mason:2010yk,Bullimore:2011ni}, except that in this case it acts in the adjoint representation. Thus, as a consequence of the superamplitude / 
supertwistor Wilson loop correspondence proved in~\cite{Mason:2010yk,Bullimore:2011ni}, the ratio of correlation functions involving our 
supermultiplet reproduces the square of the complete $n$-particle planar S-matrix (for arbitrary helicity configurations), divided by the 
MHV tree.

In~\cite{Belitsky:2011zm} the authors proposed that the correspondence between appropriately supersymmetrized correlation functions 
and amplitudes should be more robust than that between supersymmetric Wilson loops and
amplitudes. Our proof~\cite{Mason:2010yk,Bullimore:2011ni} of the Wilson loop / scattering amplitude integrand correspondence directly 
gives rise to the proof of the supersymmetric correlation function / scattering amplitude correspondence in this paper. We therefore see 
that the approaches are not so very different.

{\it Note added}: A space-time argument also relating the null-separation limit of correlators of supermultiplets to adjoint Wilson Loops was given in~\cite{Eden:2011ku}, which appeared on the arXiv simultaneously with the present work.


\section{Gauge Invariant Local Operators in Twistor Space}
\label{sec:operator}

Our first task is to construct the operator in twistor space that corresponds to the local, gauge invariant space-time operators~\eqref{BPS} or~\eqref{Konishi}. The details of the $R$-symmetry representation are not important at this stage, so in this section we consider a generic operator of the form $\cO(x) = {\rm Tr}(\Phi^2)$. 

We build up the twistor operator in stages. Firstly, consider a single scalar field $\Phi(x)$ in an Abelian theory (so that $\Phi$ is gauge invariant). A basic fact of twistor theory is that any such field that obeys its e.o.m. $\Box\Phi=0$ corresponds to a cohomology class $[\phi]$ in twistor space that may be represented by a (0,1)-form $\phi(Z)$ of homogeneity $-2$ under under a rescaling of $Z$. Concretely, the Penrose transform states that
\be	
	\Phi(x) = \int_{\rm X} \left.\la \lambda\,{\rm d}\lambda\ra \wedge\phi (Z,\bar Z) \right|_{\rm X}
\label{Pentrans}
\ee where X is the $\CP^1$ in twistor space that corresponds to the
point $x$ in space-time, 
\be
	\mu^{A'}=-\im x^{AA'}\lambda_A
	\qquad
	\chi^a=\theta^{Aa}\lambda_A
\ee
(although we are ignoring the supersymmetric coordinates in this section) and
$\lambda_A$ is taken as a homogeneous coordinate along X. As a simple
example, if 
\be
	\phi(Z,\bar Z) = \frac{1}{A\cdot Z}\,\delbar\left(\frac{1}{B\cdot Z}\right) 
\label{basicexample}
\ee 
then setting $\mu^{A'}=-\im x^{AA'}\lambda_A$ one finds that $\Phi(x) \propto 1/(x-y)^2$ on
space-time, where $y$ corresponds to the $\CP^1$ in twistor space
given by the intersection of the planes $A\!\cdot\! Z=0$ and
$B\!\cdot\! Z=0$. (See {\it e.g.}~\cite{Penrose:1986ca} for a discussion of~\eqref{basicexample} using contour integrals.) Thus, in an Abelian theory\footnote{If the individual fields in $\cO(x)$ obey their equations of motion, their corresponding twistor fields 
should represent elements of the appropriate cohomology class, {\it i.e.}, $[\phi]\in H^1(\PT',\cO(-2))$ for scalars. To consider off-shell 
fields in twistor space, we merely relax the condition $\delbar\phi=0$ that is the (linearized) equation of motion for $\phi$ following from 
the twistor action~\eqref{action}. Off-shell, $\phi$ is simply a smooth $(0,1)$-form.}
\be
	\Phi^2(x) = \int\limits_{{\rm X}\times{\rm X}}\la\lambda\,{\rm d}\lambda\ra\wedge\la \lambda'{\rm d}\lambda'\ra
	\wedge\phi(Z)\wedge\phi(Z')
\label{AbPhi2}
\ee
where $Z$ and $Z'$ are each restricted to the {\it same} Riemann sphere X, but their locations along X are integrated over separately.

The key point here is that the local space-time operator corresponds to a non-local operator on twistor space. Because of this, if $\Phi$ is in the adjoint representation of a non-Abelian group, we cannot construct the twistor operator simply by taking the trace of~\eqref{AbPhi2}.

In twistor space, non-Abelian gauge fields are described by a complex vector bundle $E\to\CP^3$ that has vanishing first Chern class. Such an $E$ is topologically trivial on restriction to any $\CP^1$ -- a necessary condition if we wish to describe a space-time theory where X corresponds to a point.  Thus one can (generically\footnote{This holomorphic triviality of the bundle is generic and will always hold in perturbation theory around the trivial bundle when $a$ has been assumed small.}) find a holomorphic frame for $E|_{\rm X}$; that is, a gauge transform $h(x,\lambda,\bar\lambda)$ on X such that
\be
	\left.h^{-1}\circ(\delbar + a)\right|_{\rm X}\circ h = \left.\delbar\right|_{\rm X}
\label{hdef}
\ee
where $\delbar+a$ is the covariant d-bar operator on $E$.

To compare the adjoint-valued field $\phi$ at $Z\in {\rm X}$ with that at $Z'\in{\rm X}$, we work in the trivialisation of $E|_{\rm X}$ defined by $h$, so that 
\be
	 {\rm Tr}(\Phi^2)(x)= \int\limits_{{\rm X}\times{\rm X}} 
	\la\lambda\,{\rm d}\lambda\ra\wedge\la \lambda'{\rm d}\lambda'\ra\wedge
	{\rm Tr}\left( h^{-1}\phi h \wedge h^{-1}\phi  h \right)\,,
\label{nonAbPhi2}
\ee
where the first factor of $h^{-1}\phi h$ is evaluated at $\lambda$ and the second at $\lambda'$. This use of holomorphic frames was an important ingredient in the construction of the twistor action for $\cN=4$ SYM in~\cite{Mason:2005zm,Boels:2006ir}, and for individual fields is the standard extension of the Penrose transform to fields in a non-trivial representation of a gauge group~\cite{HuggettTod,Hitchin:1980hp}.

\begin{figure}[t]
\begin{centering}	
	\includegraphics[width=50mm]{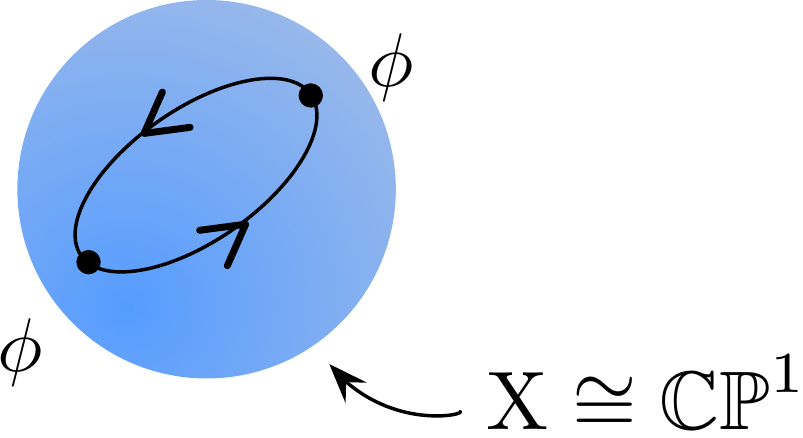}
	\caption{In a non-Abelian theory, the twistor space form of the local space-time operator ${\rm Tr}\,\Phi^2$ involves 
	holomorphic Wilson lines (or covariant propagators) on the Riemann sphere X. (The lines are intended to indicate
	only the colour flow; the propagation between the points is non-local on the sphere and not given by propagation 
	along real curves.)}
	\label{fig:nonAbPhi2}
\end{centering}
\end{figure}

It follows from~\eqref{hdef} that $h(x,\lambda,\bar\lambda)$ itself obeys $\delbar h = -a h$ on X, and we introduce the notation
\be
	U_{\rm X}(\lambda,\lambda') \equiv h(x,\lambda) h^{-1}(x,\lambda')
\label{Udef}
\ee
for the unique solution of this equation that obeys the boundary condition 
\be
	U_{\rm X}(\lambda,\lambda')|_{\lambda=\lambda'} = {\mathbb I}\ .
\ee
This normalised holomorphic frame is exactly the same object that appears in the (as yet, non-supersymmetric) twistor Wilson loop in~\cite{Mason:2010yk,Bullimore:2011ni}. As explained in those papers,  $U_{\rm X}$ can be written in terms of the twistor field $a$ as the Born series
\be
	U_{\rm X}(\lambda,\lambda') = \frac{1}{1+\delbar^{-1} a} 
	= {\rm P } \exp\left(-\int_{\rm X}\omega_{\lambda,\lambda'}\wedge a\right)\,.
\label{Ufield}
\ee
(In the first of these expressions, the inverse $\delbar$ operators
are understood to act on everything to their right. In the second,
these Green's functions are written as a meromorphic
differential $\omega$ 
on X. See~\cite{Mason:2010yk,Bullimore:2011ni}
for further details.) Using this,  equation~\eqref{nonAbPhi2} becomes 
\be
	{\rm Tr}(\Phi^2)(x) = \int\limits_{{\rm X}\times{\rm X}} 
	\la\lambda\,{\rm d}\lambda\ra\wedge\la \lambda'{\rm d}\lambda'\ra\wedge
	{\rm Tr}\left( \phi(Z)\,U_{\rm X}(\lambda,\lambda')\, \phi(Z')\, U_{\rm X}(\lambda',\lambda) \right) 
\label{2nonAbPhi2}
\ee
which may be interpreted as two insertions of the field $\phi$ connected together by a pair of holomorphic Wilson lines on X, each in the fundamental representation (see figure~\ref{fig:nonAbPhi2}). In section~\ref{sec:proof}, we will use the above twistor expression to prove the conjecture~\eqref{conj} to all loops at the level of the integrand.


\section{An Operator Supermultiplet on Twistor Space}
\label{sec:super}

Before proceeding to the proof, in this section we first introduce a straightforward supersymmetric generalisation. For definiteness we 
consider the case of the Konishi operator
\be	
	\cO_{\cK} = {\rm Tr}(\overline\Phi^{ab}\Phi_{ab})\ .
\label{Konishi}
\ee
The twistor operator equivalent to $\cO_{\cK}$ was constructed using the (0,1)-form $a(Z,\bar Z)$ (as well as two powers of $\phi_{ab}$). From the perspective of $\cN=4$ SYM on twistor space, $a$ is just the lowest component of the superfield $\cA(Z,\bar Z,\chi)$ of~\eqref{supermultiplet}, suggesting that we should instead consider the operator 
\be
	\cO(x,\theta) \equiv \epsilon^{abcd}\int\limits_{{\rm X}\times{\rm X}}
	\la\lambda\,{\rm d}\lambda\ra\,\la \lambda'{\rm d}\lambda'\ra\ 
	{\rm Tr}\left( \frac{\del^2\!\cA}{\del\chi^a\del\chi^b}(Z)\,{\rm U}_{\rm X}(\lambda,\lambda')\, 
	\frac{\del^2\!\cA}{\del\chi^c\del\chi^d}(Z')\, {\rm U}_{\rm X}(\lambda',\lambda) \right) 
\label{superop}
\ee
where
\be
\frac{\del^2\!\cA}{\del\chi^a\del\chi^b} = \phi_{ab}(Z,\bar Z) + \epsilon_{abcd}\chi^c\tilde\gamma^d(Z,\bar Z)
	+ \frac{1}{2!}\epsilon_{abcd}\chi^c\chi^d\, g(Z,\bar Z)
\label{twoderiv}
\ee
is the bottom half of the supermultiplet. The operator
\be
	{\rm U}_{\rm X}(\lambda,\lambda') = {\rm
          P}\,\exp\left(-\int_{\rm X}
          \omega_{\lambda,\lambda'} \wedge\cA\right) 
\ee
is built as in~\eqref{Ufield}, but using the {\it full twistor superfield} $\cA$. X is now interpreted as a $\CP^1$ inside $\cN=4$
supertwistor space $\CP^{3|4}$ and is associated to a point
$(x,\theta)$ in chiral super space-time.   ${\rm U}_{\rm X}$ is thus a
supersymmetric version of the holomorphic Wilson line along X. Exactly
this U$_{\rm X}$ was used in~\cite{Mason:2010yk,Bullimore:2011ni} to
obtain the supersymmetric Wilson loop in the fundamental representation in twistor space that is dual to the full planar amplitude, not just 
the MHV sector. We shall see that it plays a similar role here.

The rest of this section is devoted to proving that the operator~\eqref{superop} resulting from the na{\" i}ve replacement $a\mapsto\cA$ is 
indeed a well-defined, gauge invariant operator on twistor space. In fact, we shall see that it is the twistor form of the chiral half 
({\it i.e.}, $\bar\theta=0$) of the Konishi supermultiplet. It may seem surprising that we will be able to prove the 
correspondence~\eqref{conj} using the Konishi operator rather than the protected BPS operator. In particular, $\cO_{\cK}$ has 
anomalous dimensions which may be expected to provide further divergences, not balanced by the free-field correlator 
$G^{(0)}(x_1,\ldots,x_n)$ in the null separation limit. Indeed, if one attempts to construct the correspondence~\eqref{conj} at the level of 
the integrated objects then, as discussed in~\cite{Alday:2010zy}, one must account for these additional, sub-leading divergences (whose 
detailed form depends on the exact field content of the non-protected operators). However, we shall show in the following section that, if 
one compares only the integrands and takes the null separation limit at this level, then no such correction factors are necessary. Of course, we could equally consider the 
twistor space form of the chiral part of the supercurrent multiplet containing~\eqref{BPS}, but we think it is interesting to see that the 
correspondence~\eqref{conj} holds for a much wider class of operators, if taken at the level of the integrand.

\medskip

To check that~\eqref{superop} is just the chiral part of the Konishi multiplet, begin by recalling 
from~\cite{Mason:2010yk,Bullimore:2011ni} that ${\rm U}_{\rm X}(\lambda,\lambda')$ is defined to be the unique solution of 
\be
	\left.(\delbar +\cA)\right|_{\rm X} {\rm U}_{\rm X}=0\
\label{super-hol-frame}
\ee
on the line X in supertwistor space that is the identity when $\lambda=\lambda'$. As in~\eqref{Udef}, the supersymmetric U can be written as 
\be
	{\rm U}_{\rm X}(\lambda,\lambda') = H(x,\theta,\lambda,\bar\lambda)H^{-1}(x,\theta,\lambda',\bar\lambda')
\label{Uholframe}
\ee
in terms of an arbitrary holomorphic frame $H$ that now depends on
$\theta^{Aa}$. Because $\cA$ depends on $\theta$ only through its
dependence on $\chi$,  $\lambda^A{\del}_{Aa}\cA=0$, where
$\del_{Aa}\equiv{\del}/{\del\theta^{Aa}}$. Differentiating~\eqref{super-hol-frame}
thus gives 
\be
	\left.\left(\delbar +\cA\right)\right|_{\rm X} \left(\lambda^{A}\partial_{Aa} {\rm U}\right)=0
\ee
ensuring that ${\rm U}^{-1}(\lambda^{A}\partial_{Aa}  {\rm U})$ is
globally holomorphic on X (here and in the following we are holding $\lambda'$ constant). This expression clearly has homogeneity
$+1$ through its dependence on $\lambda$, and the only globally
holomorphic objects of homogeneity 1 are linear in $\lambda$, so we
must have  
\be
	{\rm U}^{-1}(\lambda^{A}\partial_{Aa}  {\rm U})=\lambda^{A}\,\Gamma_{Aa}(x,\theta) 
\label{conn-construction}
\ee
for some $\Gamma_{Aa}(x,\theta)$ and one finds that $D_{Aa}=\partial_{Aa}+\Gamma_{Aa}$ transforms as a connection in the $\theta$ direction. It follows from the construction that $\lambda^{A}D_{Aa}\,{\rm U}^{-1}=0$, so this connection satisfies the integrability condition
\be
	\{D_{a(A}\, ,D_{B)b}\}=0\, .
\label{integrability}
\ee
Hence the only nontrivial part of its curvature in the fermionic directions is $\cW_{ab}\,{\rm d}\theta^a_{A}{\rm d}\theta^{Ab}$, where
\be
	\cW_{ab}
	= \del_{A[a}\Gamma_{b]}^{A}+ \left\{\Gamma_{\ a}^{A}\, , \Gamma_{Ab}^{\phantom{A}}\right\}
\label{odd-curv}
\ee
is a Lorentz singlet, antisymmetric in the $R$-symmetry indices. This connection in the $\theta$ directions and
corresponding curvature (together with the integrability conditions \eqref{integrability}) are
perhaps best understood as the odd part of the $\bar\theta=0$ part of the space-time superconnection~\cite{Witten:1978xx,Harnad:1988rs}. However, a key point is that, unlike the bosonic part of the superconnection in~\cite{Mason:2010yk}, to obtain $\cW_{ab}$ we did not need to impose any part of the twistor space field equations.

The formula for $\cW_{ab}$ simplifies considerably once we realise that, since U is the identity when $\lambda=\lambda'$, equation~\eqref{conn-construction} implies ${\lambda'}^{A}\Gamma_{Aa}=0$ so that $\Gamma_{Aa}= \lambda'_A\Gamma_a$ for some fermionic scalar $\Gamma_a(x,\theta)$. Therefore, with our choice of initial condition for the holomorphic frame U, the second term in~\eqref{odd-curv} vanishes and
$\cW_{ab} = \del_{A[a}\Gamma^A_{\ b]}={\lambda'}^A\del_{A[a}\Gamma_{b]}$.

We now obtain a formula for $\Gamma_a(x,\theta)$ and
$\cW_{ab}(x,\theta)$ directly in terms of the twistor field
$\cA$. Since U always obeys~\eqref{super-hol-frame}, we have
\be	
	\int\limits_{\rm X} \frac{\la \lambda''\lambda'\ra\,\la\lambda\,\rd\lambda\ra}
	{\la\lambda''\lambda\ra\ \la\lambda\lambda'\ra}
	\ {\rm U}(\lambda'',\lambda)\left(\delbar+\cA\right){\rm
          U}(\lambda,\lambda')=0\, ,
\ee
where, $\cA$ is evaluated at $\lambda\in{\rm X}$.  Differentiating
with respect to $\theta$ and integrating the $\delbar( \del {\rm
  U}/\del\theta)$ term by parts and using the fact that $\cA$ depends
on $\theta$ only through $\chi$, one finds
\be
	\frac{\del {\rm U}(\lambda'',\lambda')}{\del\theta^{Aa}} = \int\frac{\la \lambda''\lambda'\ra\,\la\lambda\,\rd\lambda\ra}
	{\la\lambda''\lambda\ra\ \la\lambda\lambda'\ra}\ 
	{\rm U}(\lambda'',\lambda) \left(\lambda_A\frac{\del\cA}{\del\chi^a}\right) {\rm U}(\lambda,\lambda')\, ,
\ee
and so from~\eqref{conn-construction} we have
\be
\begin{aligned}
	\Gamma_a(x,\theta) &= \frac{1}{\la\lambda''\lambda'\ra} 
	{\rm U}^{-1}(\lambda'',\lambda')\,{\lambda''}^A\frac{\del{\rm U}(\lambda'',\lambda')}{\del\theta^{Aa}}\\
	&=\int\frac{\la\lambda\,\rd\lambda\ra}{\la\lambda\,\lambda'\ra}
	\,{\rm U}(\lambda',\lambda)\,\frac{\del\cA}{\del\chi^a}\,{\rm U}(\lambda,\lambda')\, .
\end{aligned}
\ee	
Using the facts that $\cW_{ab} = {\lambda'}^A\del_{A[a}\Gamma_{b]}$ and that ${\lambda'}^A\del_{Aa}{\rm U}(\lambda,\lambda')=0$, one readily finds that the odd-odd supercurvature is
\be
	\cW_{ab}=\partial^{A}_{a}\Gamma_{Ab}
	= -\int\limits_{\rm X}\la\lambda\,\rd\lambda\ra \wedge 
	{\rm U}(\lambda',\lambda) \frac{\del^2\!\cA}{\del\chi^a\del\chi^b}{\rm U}(\lambda,\lambda')\, 
\ee
when expressed on twistor space. Expanding to second order in the fermionic components using equation~\eqref{twoderiv} we find 
\be
\cW_{ab} = \Phi_{ab} + \epsilon_{abcd}\theta^{cA}(\tilde{\Psi}_A^d + \frac{1}{2!}\theta^{dB}G_{AB}) - [\Phi_{a(c},\Phi_{d)b}]\theta^{cA}\theta^{d}_{A}+ O(\theta^3).
\label{fermcurv}
\ee
The integrability conditions~\eqref{integrability} ensure that only space-time fields $\{\Phi_{ab},\tilde{\Psi}^a_A,G_{AB}\}$ appear in the expansion. Therefore this operator contains the chiral half of the $\cN=4$ vector multiplet.

We are now in position to interpret the twistor operator~\eqref{superop} more invariantly. Using the concatenation property ${\rm U}(\lambda_1,\lambda_2){\rm U}(\lambda_2,\lambda_3) = {\rm U}(\lambda_1,\lambda_3)$ that follows from~\eqref{Uholframe}, we have\footnote{In Lorentzian signature, one imposes the reality condition $\epsilon^{abcd}\cW_{cd} = g^{a\bar c} g^{b\bar d}\overline{\cW}_{\bar c\bar d}$, where $g_{a\bar b}$ is an Hermitian metric that preserves an SU(4) subgroup of the complexified $R$-symmetry group SL($4;\mathbb{C})$. Since the integrand is a rational function, there is no need to impose any reality condition on the multiplet at this level.}
\be
\begin{aligned}
	\cO(x,\theta)&\equiv \epsilon^{abcd}\,{\rm Tr} \,\cW_{ab}\cW_{cd} \\
	&= \epsilon^{abcd}\int\limits_{{\rm X}\times{\rm X}}\la\lambda\rd\lambda\ra\la\lambda'\rd\lambda'\ra\wedge
	{\rm Tr} \left[{\rm U}(\lambda,\lambda')
          \frac{\del^2\!\cA(\lambda')}{\del\chi^a\del\chi^b}\,
          {\rm U}(\lambda',\lambda)\frac{\del^2\!\cA(\lambda)}{\del\chi^c\del\chi^d}\right]\,   , 
\end{aligned}
\label{superKon} 
\ee
so that~\eqref{superop} is the twistor form of the trace of the square of the odd-odd curvature on chiral super space-time. This operator is by construction invariant under both gauge transformations and the chiral half ($Q$ and $\bar S$) of the superconformal algebra. Since our construction did not impose any field equations, these properties hold even off-shell. Using the component expansion of the fermionic curvature \eqref{fermcurv}, we find 
\begin{multline}
\cO(x,\theta) = \mathrm{Tr} \left(\overline{\Phi}^{ab} \Phi_{ab}\right) + 2\epsilon_{abcd}\theta^{cA}\mathrm{Tr}\left(\overline{\Phi}^{ab}\tilde{\Psi}^{d}_{A}\right)+\epsilon_{abcd}\theta^{cA}\theta^{dB}\mathrm{Tr}\left(\overline{\Phi}^{ab}G_{AB}\right) \\
-\frac{1}{2}\theta^{cA}\theta^{d}_{A}\mathrm{Tr}\left(\overline{\Phi}^{ab}[\Phi_{a(c},\Phi_{d)b}]\right)-4\epsilon_{abcd}\theta^{aA}\theta^{cB}\mathrm{Tr}\left(\tilde{\Psi}^{b}_{A}\tilde{\Psi}^{d}_{B}\right)+O(\theta^3),
\end{multline}
and one may identify $\cO(x,\theta)$ with the usual Konishi supermultiplet at $\bar\theta=0$. This chiral part of the multiplet is not invariant under off-shell anti-chiral supersymmetry transformations ($\bar Q$ and $S$) -- indeed, repeatedly acting on ${\rm Tr}\, \cW_{ab}\cW^{ab}$ with $\bar Q$s would fill out the full Konishi supermultiplet. However, in studying the correspondence at the level of the integrand, only the chiral part plays a role.


\section{Proving the Correspondence}
\label{sec:proof}

\begin{figure}[t]
\begin{centering}
	\includegraphics[width=100mm]{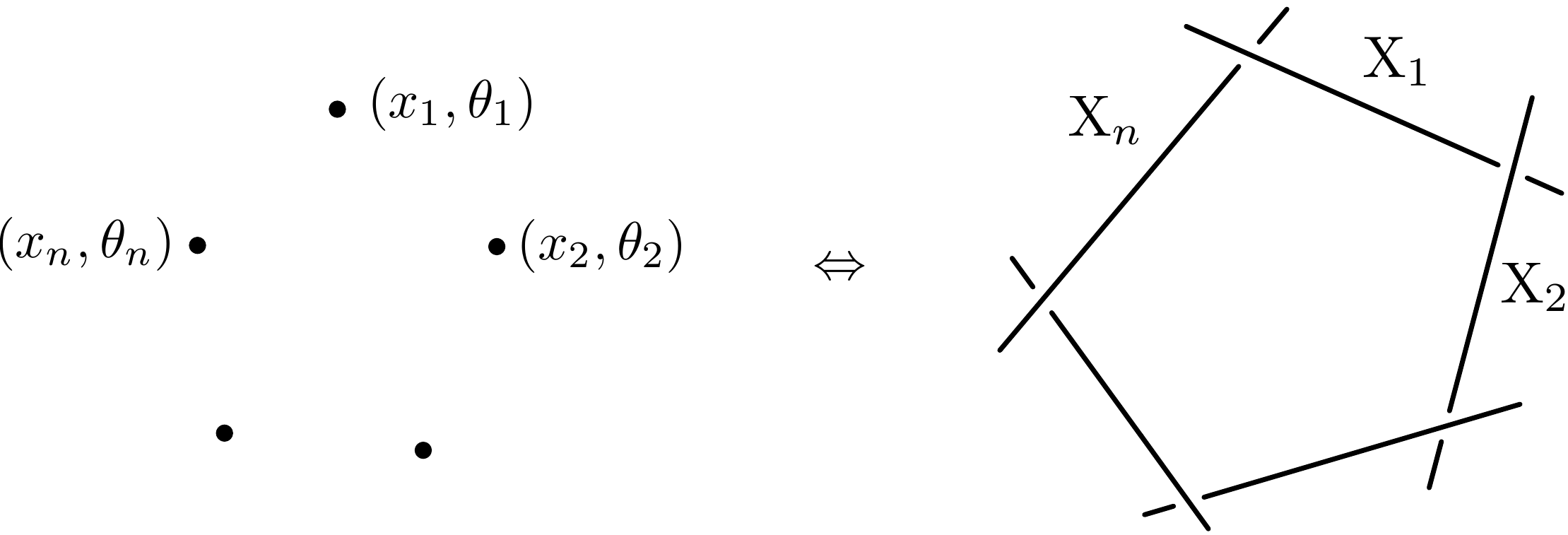}
	\caption{The $n$ generic points $(x,\theta)$ correspond to $n$ lines ($\CP^1$s) in $\CP^{3|4}$. In the limit that 
	$(x_i-x_{i+1})\to\tilde\lambda_i\lambda_i$ and $\theta_i-\theta_{i+1} \to \eta_i\lambda_i$, these lines intersect in 
	supertwistor space, forming a nodal curve $C$.}
	\label{fig:polygons}
\end{centering}
\end{figure}

In this section, we prove (the supersymmetric version of) the conjectures of~\cite{Alday:2010zy,Eden:2010zz,Eden:2010ce}. More precisely, the statement we shall prove is the following:

In the limit that both $x_{i,i+1}^2\to 0$ (so that $(x_i-x_{i+1})\to\tilde\lambda_i\lambda_i$ for some null vector $\tilde\lambda_i\lambda_i$) and $(\theta_i^{Aa}-\theta_{i+1}^{Aa})\lambda_{iA} \to0$, the ratio of $n$-point correlators
\be
	\frac{\cG(x_1,\theta_1;\ldots;x_n,\theta_n)}{G^{(0)}(x_1,\ldots,x_n)} \equiv 
	\frac{\la 0| \cO(x_1,\theta_1)\cdots \,\cO(x_n,\theta_n)|0\ra}{\la 0|\cO(x_1)\cdots\, \cO(x_n)|0\ra_{\rm tree}}
\label{superratio}
\ee
is equal at the level of the integrand to the correlator 
\be
\frac{1}{N^2-1}\big\langle\; \mathrm{Tr}_{\mathrm{Adj}}\; \mathrm{P} \exp\left( - \int_C \omega \wedge \cA \right)\;\big\rangle
\label{sWL}
\ee
of the supersymmetric twistor Wilson loop, in the adjoint representation, around the curve $C\in \CP^{3|4}$ that corresponds to the null polygon $\gamma$ in chiral super space-time (see figure~\ref{fig:polygons}). 

By `{\it at the level of the integrand}' we mean that we allow all possible Feynman diagrams (to all loops, planar and non-planar) coming from contracting the operators with arbitrarily many vertices from the action, but we {\it do not perform the integrals over the locations of these vertices}. As in~\cite{Eden:2010ce}, we assume that all such vertices are at generic locations and in particular that they are not null separated from any of the insertion points $x_i$. 

Note that the denominator of~\eqref{superratio} involves the tree-level correlation function of the non-supersymmetric operator ${\rm Tr}\,\Phi_{ab}\Phi^{ab} = \left.{\rm Tr}\, \cW_{ab}\cW^{ab}\right|_{\theta=0}$ only. For arbitrary insertion points, this term is given by
\be
	G^{(0)}(x_1,\ldots,x_n) = \frac{m}{x_{12}^2 x_{23}^2\cdots x_{n1}^2} + \hbox{other permutations}\,,
\label{treecorr}
\ee
where $m$ is a numerical constant that cancels in the ratio. The displayed term is the leading term as $x_{i,i+1}^2\to0$ for all $i$. Its presence in the denominator of~\eqref{superratio} means that only those numerator terms that are (at least) as divergent survive in the integrand when the limiting configuration is reached.

We now consider the numerator. While it is perfectly possible to proceed on space-time using~\eqref{superKon}, it is more convenient to work with the twistor form~\eqref{superop}. This equation expresses $\cO(x,\theta)$ in terms of twistor fields, so to compute its correlation function we must use the twistor action of $\cN=4$ SYM~\cite{Boels:2006ir}
\be
	S[\cA] = \int\limits_{\CP^{3|4}}\hspace{-0.2cm}
	{\rm D}^{3|4}Z \wedge {\rm Tr}\left(\frac{1}{2}\cA\,\delbar\cA + \frac{1}{3}\cA^3\right)
	+{\rm g}^2\int\limits_\M \rd^{4|8}x\,\log\det\left(\delbar+\cA\right)_{\rm X}
\label{action}
\ee
that contains a holomorphic Chern-Simons theory plus an infinite sum
of MHV vertices, each supported on some  ${\rm X}\subset\CP^{3|4}$
corresponding to the location of the vertex in space-time. This sum of MHV vertices is equivalent under a gauge transform on twistor 
space to the non self-dual part of the space-time $\cN=4$ Lagrangian  of Chalmers \& Siegel~\cite{Chalmers:1996rq} and are the chiral 
Lagrangian insertions discussed in~\cite{Eden:2010ce}. Eventually,the location of these vertices should be integrated over along some
choice of real slice $\M$ of complexified space-time\footnote{The Grassmann integration is always performed algebraically.}, but in 
studying the integrand we wish to examine the behaviour of Feynman diagrams before carrying out these integrals. The genericity 
assumption on the space-time interaction vertices corresponds to the assumption that these MHV vertex lines do not intersect the lines 
X$_i$ on which the operators $\cO(x_i,\theta_i)$ are supported.

Contractions between fields, either in the operator insertions or MHV vertices, are performed using the axial gauge propagator
\be
	\la \cA(Z)^I_{\ J}\,\cA(Z')^K_{\ L}\ra 
	= \bar\delta^{2|4}(Z_*,Z,Z') \left(\delta^I_{\ L} \delta^K_{\ J} - \frac{1}{N}\delta^I_{\ J}
	\delta^K_{\ L}\right)\,,
\label{superprop}
\ee
for the SU$(N)$ holomorphic Chern-Simons theory on $\CP^{3|4}$, where $Z_*$ is a reference twistor that defines the axial gauge. We take $Z_*$ to be generic ({\it i.e.}, we choose $Z_*$ not to lie on any of the operator insertion lines X$_i$). The projective delta-function 
\be
	\bar\delta^{2|4}(Z_*,Z,Z')\equiv\int\frac{\rd s}{s}\frac{\rd t}{t}\,\bar\delta^{4|4}(Z_*+sZ+tZ')
\label{proj-delta}
\ee
is superconformally invariant and has support where its three arguments are collinear in the projective space. (See~\cite{Mason:2009sa,Adamo:2011cb} for an introduction to these projective delta-functions.)  

To obtain a non-zero correlator (even away from the limit), each of the $2n$ explicit powers of $\del^2\!\cA/\del\chi^2$ in the 
product of the $n$ twistor operators~\eqref{superop} must be contracted. All contractions can occur either directly between fields on 
different\footnote{The implicit normal ordering of composite operators means that all contractions occur between fields inserted on 
distinct lines in twistor space ({\it i.e.}, there are no contractions between fields on the same line).} lines X$_i$ or via MHV vertices, which 
are the only vertices remaining in the twistor action in the axial gauge.  The fields in the operator insertions include both the explicit powers of the $\del^2\!\cA/\del\chi^2$ and arbitrarily many further powers of $\cA$ from the expansion of the ${\rm U}_{\rm X}$s. There are three classes of contraction to consider:
\begin{enumerate}
\item Contractions between the operator insertion on some line X$_i$ and fields in a (MHV) vertex from the Lagrangian,

\item Contractions between the operator insertions on non-consecutive lines X$_i$ and X$_j$.

\item Contractions between the operator insertions on consecutive lines X$_i$ and X$_{i+1}$.
\end{enumerate}
We will see that the only terms which contribute to the {\it integrand} of the ratio~\eqref{superratio} in the null limit come from this final 
class, where the explicit $\del^2\cA/\del\chi^2$ insertions on consecutive lines are contracted. Note that Feynman diagrams that 
contribute to the anomalous dimensions of operators such as the Konishi are a subset of the first class.

In what follows, we parameterize the line X$_i$ by $Z_{A_i}+s_i Z_{B_i}$ so that $s_i$ is an inhomogeneous coordinate on 
X$_i$. The measure $\langle\lambda_i \rd\lambda_i\rangle$ then becomes $\langle A_{i}B_{i}\rangle \rd s_i$.  It will also be 
useful to recall the definition of the `R-invariant' --  the standard chiral invariant of the superconformal group -- which depends on  
five arbitrary twistors:
\be
\begin{aligned}
	[1,2,3,4,5]
	&:=\int\frac{\rd^4s}{s_1s_2s_3s_4}\,\bar{\delta}^{4|4}\!\left(Z_1+\sum_{i=1}^4s_iZ_i\right) \\
	&=\frac{\delta^{0|4}\!\left( (1234)\chi_5 + \mbox{cyclic}\right)}{(1234)(2345)(3451)(4512)(5123) }
\label{superconf}
\end{aligned}
\ee
where $(1234)=\epsilon_{\alpha\beta\gamma\delta}Z_1^\alpha Z_2^\beta
Z_3^\gamma Z_4^\delta$ {\it etc.}, and the second line is obtained by integration against the bosonic delta functions, see \cite{Mason:2009qx} for details.

\medskip 

Consider first a contraction between a field inside a holomorphic frame U$_{\rm X_i}$ in one of the operator insertions and a field $\cA$ 
in an MHV vertex supported on some line X=$C+tD$. Exactly the same calculation as in~\cite{Mason:2010yk} shows that this contraction is
\be
	\int \frac{\rd s}{s}\frac{\rd t}{t}\, \bar\delta^{2|4}(Z(s),Z_*,Z(t))  = [A_i,B_i,*,C,D]
\ee
Similarly, a contraction between an explicit power of $\del^2\!\cA/\del\chi^a\del\chi^b$ in the operator on the line X$_i$ and a field $\cA$ 
from MHV vertex yields 
\begin{multline}
	\frac{\del^2}{\del\chi_{A_i}\del\chi_{B_i}} [A_i,B_i, *,C,D]\\
	= \frac{\delta^{0|2}\!\left(\chi_{A_i}(B_i*ZB_{i+1})\! +\! \chi_{B_i}(*B_iB_{i+1}A_i)\!+\!\chi_*(ZB_{i+1}A_iB_i)\!+\!\chi(B_{i+1}A_iB_i*)
	\!+\!\chi_{B_{i+1}}(A_iB_i*Z)\right)}{(A_iB_iCD)(DA_iB_i*)(A_iB_i*C)}\, .
\label{2diffR}
\end{multline}
Our genericity assumption guarantees that ${\mathrm{X}}\cap \mathrm{X}_i=\emptyset$ even in the null limit, so none of the factors in the 
denominator of these R-invariants vanish. Therefore, contractions of this type do not provide a divergence in the integrand to balance the 
denominator of~\eqref{superratio}. Here it is important that we are considering taking the null separation limit {\it already at the level of the 
integrand}: Feynman diagrams involving MHV vertices may well lead to divergences once the location of this MHV vertex is integrated 
over -- leading among other things to anomalous dimensions and operator mixing in the interacting theory -- but they cannot supply the 
required divergence at the level of the integrand.

In the same way, contractions between fields on non-adjacent insertion lines lead to the R-invariant $[A_i,B_i, *, A_j, B_j]$ or derivatives 
thereof, depending on whether we contract fields in the holomorphic frames or the explicit $\del^2\!\cA/\del\chi^2$s. Since every four-
bracket in this R-invariant remains non-vanishing as we take null separated limit between {\it adjacent} insertions, this class of 
contractions also remains finite. Thus, considering only contractions of the first two classes cannot lead to a divergence of the integrand 
that balances the denominator of~\eqref{superratio}.

It is not surprising that, for the integrand to diverge, we must contract fields on adjacent lines. However, here there is a delicate point: 
the free-field correlator in the denominator~\eqref{superratio} only knows about contractions between scalar fields at adjacent space-time 
points, whereas in the numerator, adjacent contractions could occur either between pairs of holomorphic frames, or pairs of $\del^2\!\cA/\del\chi^2$s, or between a holomorphic frame and a $\del^2\!\cA/\del\chi^2$. Let us consider each case in turn, and carefully examine their behaviour as X$_i$ and X$_{i+1}$ are made to intersect.

Firstly, contractions between two holomorphic frames on gives $[A_i,B_i,*,A_{i+1},B_{i+1}]$. To study the behaviour of this object in the 
null separated limit, suppose $A_{i+1} = B_i + \epsilon Z$ for some twistor $Z$. In the limit $\epsilon\to0$ one finds\footnote{To obtain this 
result, it is of course important that one sets $A_{i+1}=B_i + \epsilon Z$ as a relation on the {\it super}twistors, ensuring that the lines 
intersect in $\CP^{3|4}$.}
\begin{multline}
	\lim_{\epsilon\to0} \ [A_i,B_i,*,(B_i + \epsilon Z),B_{i+1}] \\
	= \epsilon\  \frac{\delta^{0|4}
	\left(\chi_{A_i}(B_i*ZB_{i+1})\! +\! \chi_{B_i}(*B_iB_{i+1}A_i)\!+\!\chi_*(ZB_{i+1}A_iB_i)\!+\!\chi(B_{i+1}A_iB_i*)
	\!+\!\chi_{B_{i+1}}(A_iB_i*Z)\right)}{(A_iB_i*Z)(B_i*ZB_{i+1})(*B_iB_{i+1}A_i)(ZB_{i+1}A_iB_i)(B_{i+1}A_iB_i*)}
\end{multline}
The overall factor of $\epsilon$ comes from the ratio of $\epsilon^4$ from the fermionic numerator to $\epsilon^3$ from the bosonic 
denominator. Therefore such contractions actually vanish in the null limit due to cancellation between the supermultiplet.

The only difference when considering contractions between the holomorphic frame and a $\del^2\!\cA/\del\chi^2$ insertion, or between 
two $\del^2\!\cA/\del\chi^2$ insertions on adjacent lines is that we must differentiate the R-invariant with respect to the fermions before 
taking the $\epsilon\to0$ limit. For the contraction involving a single holomorphic frame, from~\eqref{2diffR} we have
\begin{multline}
	\left.\lim_{\epsilon\to0} \left[\frac{\del^2}{\del\chi_{A_i}\del\chi_{B_i}} [A_i,B_i, *,A_{i+1},B_{i+1}]\right|_{A_{i+1}=B_i+\epsilon Z}\right]
	\\
	= \frac{\delta^{0|2}\!\left(\chi_{A_i}(B_i*ZB_{i+1})\! +\! \chi_{B_i}(*B_iB_{i+1}A_i)\!+\!\chi_*(ZB_{i+1}A_iB_i)\!+\!\chi(B_{i+1}A_iB_i*)
	\!+\!\chi_{B_{i+1}}(A_iB_i*Z)\right)}{(A_iB_iZB_{i+1})(B_{i+1}A_iB_i*)(A_iB_i*Z)}
\label{2diffRlim}
\end{multline}
which, while non-vanishing, remain finite in the null separated limit, and depend on the direction in which null separation was approached.

Equations~\eqref{2diffR} and~\eqref{2diffRlim} show that we cannot hope to balance the denominator of~\eqref{superratio} if we contract 
the explicit insertions of $\del^2\!\cA/\del\chi^2$ either with fields in an MHV vertex from the Lagrangian, or with holomorphic frames, even 
on an adjacent line. Nor can we balance the denominator using contractions between pairs of holomorphic frames. Thus, the only hope 
to obtain a non-vanishing result if null separation is taken at the level of the integrand is to have contracted the explicit 
$\del^2\!\cA/\del\chi^2$ insertions with eachother on adjacent lines, without any intermediate MHV vertices or holomorphic frames. To see that this does provide the correct divergence, note that
\be
\begin{aligned}
	&\la A_{i} B_{i}\ra \la A_{i+1}B_{i+1}\ra 
	\int\limits_{\rm{X}_{i}\times\rm{X}_{i+1}}\la\lambda_i \rd\lambda_{i}\ra\la\lambda_{i+1}\rd\lambda_{i+1}\ra\, 
	\left\la\frac{\del^{2}\cA}{\del\chi^{a}\del\chi^{b}}(Z(\lambda_i))\,\frac{\del^{2}\cA}{\del\chi^{c}\del\chi^{d}}(Z(\lambda_{i+1}))\right\ra\\
	=\ &\langle A_{i} B_{i}\rangle\langle A_{i+1}B_{i+1}\rangle\int \rd s_{i}\,\rd s_{i+1}\,
	\bar{\delta}^2(A_{i}+s_{i}B_{i},Z_*,A_{i+1}+s_{i+1}B_{i+1}) \\
	=\ &\epsilon_{abcd}\frac{\la A_{i} B_{i}\ra\la A_{i+1}B_{i+1}\ra}{(A_iB_iA_{i+1}B_{i+1})} 
	= \frac{\epsilon_{abcd}}{(x_i-x_{i+1})^2}\ ,
\end{aligned}
\label{divcontract}
\ee
exactly as for the free correlator.

\medskip

\begin{figure}[t]
\begin{centering}   
    \includegraphics[width=150mm]{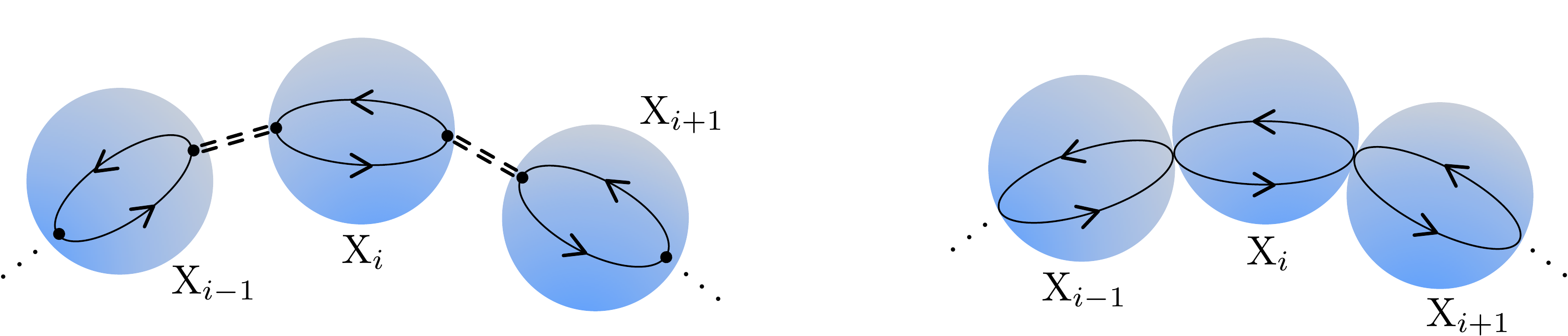}
    \caption{The only non-vanishing contribution to the integrand ratio in the null limit comes from
    direct contractions between $\phi$s on adjacent Riemann spheres. The twistor propagators freeze the
     locations of the $\phi$s on each X$_i$. In the limit that the lines intersect, these locations are the intersection
     points. The remaining operator is the supersymmetric twistor Wilson loop, acting in the adjoint.}
    \label{fig:limit}
\end{centering}
\end{figure}

This establishes that the only terms that survive in~\eqref{superratio} when the null separation limit is taken at the level of the integrand 
involve contractions between each of the $\del^2\!\cA/\del\chi_{i}^a\del\chi_{i}^b$ insertions with its neighbours on adjacent twistor lines. 
Once these explicit insertions have been contracted to balance the divergence in the denominator, the remaining fields in the 
holomorphic frames may contract in any way they wish -- either with other holomorphic frames or using the MHV vertices in the 
interacting theory. We must now show that all these remaining contractions are equivalent to the integrand of the supersymmetric Wilson 
loop in the adjoint.

To do so, note that the integrals in~\eqref{divcontract} were completely frozen by the two bosonic delta functions. Geometrically, this is the 
statement that, given a generic point $Z_*\in\CP^3$, there is a unique line L through $Z_*$ that intersects both X$_i$ and 
X$_{i+1}$. Explicitly, L is the line through $Z_*$ and the intersection of the plane $(*,A_i,B_i)$ and the line $(A_{i+1},B_{i+1})$.
When we contract two field insertions on X$_i$ and X$_{i+1}$, they are thus frozen to be at ${\rm L}\cap{\rm X}_i$ and 
${\rm L}\cap{\rm X}_{i+1}$, respectively. Now, clearly, in the limit that X$_i$ and X$_{i+1}$ are brought to intersect, L is simply the 
line through $Z_*$ and their intersection point ${\rm X}_i\cap{\rm X}_{i+1}$ so that the arguments of the holomorphic frames U adjacnet to 
the contracted fields now coincide (see figure~\ref{fig:limit}). The resulting product of holomorphic frames around the nodal curve 
is exactly the same supersymmetric twistor Wilson loop as found for the scattering amplitudes~\cite{Mason:2010yk,Bullimore:2011ni} 
\be
	\frac{1}{N^2-1}\left\la \mathrm{Tr}_{\rm{Adj}}\; \rm{P} \exp\left( - \int_C \omega \wedge \cA \right)\right\ra\, 
\ee
except now in the adjoint representation. Thus, all remaining integrand contributions to the limit of correlation functions~\eqref{superratio} 
are inevitably equal to the integrands of this adjoint super Wilson loop. This completes the proof.

\medskip

The above correspondence between the integrands of a ratio of correlator functions of local, gauge invariant operators and the integrand 
of an adjoint  Wilson loop is valid for an SU$(N)$ gauge group even at finite $N$. However, in the planar limit of a CPT invariant theory 
\be
	\la W_{\rm adj}[\gamma] \ra = \la W_{\rm fund}[\gamma] \ra\,\times\,\la W_{\rm anti-fund}[\gamma]\ra
	= \la W_{\rm fund}[\gamma]\ra^2 \, .
\ee
Using the proof~\cite{Bullimore:2011ni} of the correspondence between the fundamental twistor Wilson loop and the integrand of the scattering amplitude, we have as an immediately corollary that {\it in the planar limit}, the ratio of operators~\eqref{superratio} is equal to the {\it square} of the integrand of the planar scattering amplitude (divided as always by the MHV tree). As in~\cite{Mason:2010yk,Bullimore:2011ni,Caron-Huot:2010ek}, this includes all loop corrections to the integrands of all N$^k$MHV partial amplitudes\footnote{Of course, one must square the complete integrand of the planar super-amplitude -- the sum of the tree plus all-loop corrections -- and the compare the coefficients of a particular power of the 't Hooft coupling.}.

\medskip

Two special cases of this general correspondence are worthy of separate mention. First, if we return to the ratio of {\it non}-supersymmetric correlators, we obtain the adjoint twistor Wilson loop built purely from the non-supersymmetric field $a(Z,\bar Z)$. This non-supersymmetric twistor Wilson loop was shown in~\cite{Mason:2010yk} to correspond to a standard, non-supersymmetric Wilson loop in space-time around the null polygon $\gamma$. When taken in the fundamental representation, this Wilson loop computes the all-orders integrand for  MHV amplitudes only. 

Second, if we consider the ratio 
\be
	\lim	\frac{\cG^{({\rm sd})}(x_1,\theta_i;\ldots;x_n,\theta_n)}{G^{(0)}(x_1,\ldots,x_n)}
\ee
involving the supersymmetric operator, but computed in {\it self-dual} $\cN=4$ SYM, in the super-null limit $(x_i-x_{i+1})\lambda_i\to0$, $(\theta_i-\theta_{i+1})\lambda_i\to0$, we obtain the square of the $n$-particle scattering amplitude
\be
	M^{(0)} = 1 + \frac{A_{\rm NMHV}^{(0)}}{A^{(0)}_{\rm MHV}} + \frac{A^{(0)}_{{\rm N}^2{\rm MHV}}}{A^{(0)}_{\rm MHV}}
	+ \cdots+ \frac{A^{(0)}_{\overline{\rm MHV}}}{A^{(0)}_{\rm MHV}}
\ee
at {\it tree level} only. The corresponding statement in twistor space is that one computes the correlator ratio or Wilson loop using only the holomorphic Chern-Simons action. This truncation was already noted in~\cite{Mason:2010yk,Bullimore:2011ni}.


\section{Conclusions}
\label{sec:conclusions}

We conclude with a few remarks. 

{\it 1.)} The relationship between correlation functions of local, single trace operators and adjoint Wilson loops is much closer than the 
relationship between planar scattering amplitudes and fundamental Wilson loops. As we saw above,  the former correspondence is really 
valid at the {\it operator} level -- even supersymmetrically -- once understood in twistor space. By contrast, to prove the scattering 
amplitude / fundamental Wilson loop correspondence requires that one actually {\it evaluate} the planar integrand  on both sides 
independently, as was done in~\cite{Bullimore:2011ni,Arkani-Hamed:2010kv}. Similarly, the correlation function / adjoint Wilson loop 
correspondence is true even for finite rank gauge groups, while the relation to scattering amplitudes only arises in the planar limit. The 
reason for this closer relationship is of course that the operator and adjoint Wilson loop each live on the same space-time (or same 
twistor space), whereas this is not the space-time in which one performs the scattering experiment. (It is the dual conformal space-time.)

{\it 2.)} There is obviously a great deal of freedom in the choice of `basic' operator ${\rm Tr}\,\Phi^2$. This certainly includes the R-
symmetry representation of the scalars, but we could equally replace this operator by a bilinear in the gluinos (one of each helicity) or 
some more general choices of bilinear for each $\cO(x_i)$ separately. The key criterion is simply that operators at null-separated points 
can be contracted pairwise around the chain using propagators. Different choices of operator would lead to different behaviour in the 
sub-leading divergences of any regularised version of this correspondence at the level of the {\it integral}.  As explained, these sub-
leading terms vanish at the level of the unregularised {\it integrand} in the strict null limit.

{\it 3.)} In attempting to compare scattering amplitudes and
fundamental null Wilson loops at the level of their regularised {\it
  integrals}, one encounters the difficulty that because they
naturally live on different spaces, it is not immediately clear how a
regularisation scheme applied on one side of the correspondence should
translate to a regularisation scheme on the other. For example,
computing both sides in $4-2\epsilon$ dimensions, one needs to take
$\epsilon<0$ for the amplitude, but $\epsilon>0$ for the Wilson
loop~\cite{Drummond:2007aua}, while the Wilson loop equivalent of the
Higgs regularisation~\cite{Alday:2009zm} of the amplitude is currently
unknown. Perhaps the main interest of the present correspondence is
that, because the correlator and Wilson loop live on the same
space-time, the {\it same} regularisation scheme should be used for
both objects. After taking the planar limit, this provides a
definition of the regularised amplitude.


\acknowledgments

We thank Nima Arkani-Hamed, Simon Caron-Huot, Gregory Korchemsky, Emery Sokatchev and especially Juan Maldacena for helpful conversations. TA is supported by the NSF Graduate Research Fellowship 
(USA) and Balliol College. MB is supported by an STFC Postgraduate Studentship. LM was financed in part by EPSRC grant number EP/F016654. DS is supported by the Perimeter Institute for Theoretical Physics. Research at the Perimeter Institute is supported by the Government of Canada through Industry Canada and by the Province of Ontario through the Ministry of Research $\&$ Innovation.

\bibliographystyle{JHEP}
\bibliography{OpWLrefs}

\end{document}